\documentclass[10pt,twocolumn,superscriptaddress,showpacs,preprintnumbers]{revtex4}
\usepackage{epsf} 
\usepackage{amsmath,color}
\usepackage{amssymb}
\usepackage{epsfig}
\usepackage{latexsym}
\usepackage{amsfonts}
\usepackage{varioref}
\usepackage{ifthen}
\usepackage{graphicx}
\usepackage{amssymb,amsmath,amsthm}
\providecommand{\U}[1]{\protect\rule{.1in}{.1in}}
\begin{document}
\parindent 0mm 
\setlength{\parskip}{\baselineskip} 
\pagenumbering{arabic} 
\setcounter{page}{1}
\mbox{ }
\preprint{UCT-TP-291/12}
\preprint{MZ-TH/12-43}

\title{Hadronic contribution to the QED running coupling $\alpha(M_{Z}^2)$}
\author{S. Bodenstein}
\affiliation{Centre for Theoretical \& Mathematical Physics, University of
Cape Town, Rondebosch 7700, South Africa}
\author{C. A. Dominguez}
\affiliation{Centre for Theoretical \& Mathematical Physics, University of
Cape Town, Rondebosch 7700, South Africa}
\author{K. Schilcher}
\affiliation{Centre for Theoretical \& Mathematical Physics, University of
Cape Town, Rondebosch 7700, South Africa}
\affiliation{Institut f\"{u}r Physik, Johannes Gutenberg-Universit\"{a}t Mainz,
Staudingerweg 7, D-55099 Mainz, Germany}

\author{H. Spiesberger}
\affiliation{Institut f\"{u}r Physik, Johannes Gutenberg-Universit\"{a}t Mainz,
Staudingerweg 7, D-55099 Mainz, Germany}

\date{\today}
\begin{abstract}
\noindent
We introduce a model independent method for the determination of the hadronic contribution to the QED running coupling, $\Delta\alpha_{\text{HAD}}(M_{Z}^{2})$, requiring no $e^+e^-$ annihilation data as input. This is achieved by calculating the heavy-quark contributions entirely in perturbative QCD, whilst the light-quark resonance piece is determined using available lattice QCD results. Future reduction in the current uncertainties in the latter shall turn this method into a valuable alternative to the standard approach. Subsequently, we find that the precision of current determinations of $\Delta\alpha_{\text{HAD}}(M_{Z}^{2})$ can be improved by some $20\%$ by computing the heavy-quark pieces in PQCD, whilst using $e^+e^-$ data only for the low-energy light-quark sector. We obtain in this case $\Delta\alpha_{\text{HAD}}(M_{Z}^{2})=275.7(0.8)\, \times 10^{-4}$, which currently is the most precise value of $\Delta\alpha_{\text{HAD}}(M_{Z}^{2})$. 
\end{abstract}
\noindent
\maketitle
\section{INTRODUCTION}
Of the subset of three parameters that enter the electroweak sector of the Standard Model (SM) of particle physics, $G_F, M_Z$ and $\alpha(M_{Z}^{2})$, the least precisely known is the electromagnetic coupling at the $Z$ boson mass, $\alpha(M_{Z}^{2})$. This is primarily due to  hadronic contributions which are not calculable using perturbative QCD (PQCD). Increasing the precision of $\alpha(M_{Z}^{2})$ is important for, amongst other things, obtaining a Standard Model  fit of the Higgs mass. Currently, there is a minor tension between the recently measured mass of a potential Higgs boson, $M_H=  126.0 (0.4)(0.4)\,\text{GeV}$  \cite{higgsm1}, or $M_H=  125.3 (0.4)(0.5)\,\text{GeV}$ \cite{higgsm2}, and a mass of $91^{+30}_{-23}\,\text{GeV}$ (at the 68\% confidence level) obtained from global SM fits to electroweak precision data \cite{davier2011}.\\
The running QED coupling $\alpha$ can be parameterized as 
\begin{equation}
\alpha(s)=\frac{\alpha(0)}{1-\Delta\alpha_\text{L}(s)-\Delta\alpha_\text{HAD}(s)}\;,
\end{equation}
where $\Delta\alpha_\text{L}$ is the leptonic contribution, which can be determined with high precision in perturbation theory, and $\Delta\alpha_\text{HAD}$ is the hadronic term. Of particular interest is the QED coupling at the scale $M_Z$. Denoting $\alpha\equiv \alpha(0)$ in the sequel, $\Delta\alpha_{\text{HAD}}(M_{Z}^{2})$ can be written as
\begin{equation}\label{eq:alpha1}
\Delta\alpha_{\text{HAD}}(M_{Z}^{2})=4 \,\pi \, \alpha\left\{\Pi(0) - Re \,[\Pi(M_{Z}^{2})]\right\}\;,
\end{equation}
where $\Pi(s)$ is the standard electromagnetic current correlator
\begin{eqnarray}
\Pi_{\mu\nu} (q^2) &=& i \int d^4x\,  e^{iqx} \langle 0|\, T \left(j^{\text{\,EM}}_{\mu}(x), j^{\text{\,EM}}_{\nu}(0) \right)|0\rangle \nonumber\\ 
&=& (q_\mu q_\nu - q^2 g_{\mu\nu}) \Pi(q^2)\;,
\end{eqnarray}
with  $j^{\text{EM}}_\mu(x)=\sum_f Q_f \bar{f}(x) \gamma_\mu f(x)$, and the sum is over all quark flavors $f=\{u,d,s,c,b,t\}$, with charges $Q_f$. Invoking analyticity and unitarity for $\Pi(s)$, and using the optical theorem, i.e. $R(s)=12\pi\, \text{Im}\,\Pi(s)$, where $R(s)$ is the normalized $e^+e^-$ cross-section, one can write Eq.\eqref{eq:alpha1} as a dispersion integral \cite{cabibbo1961}
\begin{equation}\label{EQ:dispersion}
\Delta\alpha_{\text{HAD}}(M_{Z}^{2})=\frac{\alpha \, M_{Z}^{2}}{3\, \pi}P\int^{\infty}_{4m_{\pi}^{2}}\frac{R(s)}{s(M_{Z}^{2}-s)}ds \;,
\end{equation}
where $P$ denotes the principal part of the integral. This dispersion relation is useful as it only requires knowledge of $R(s)$, which can be determined experimentally.
The standard approach to determining $\Delta\alpha_{\text{HAD}}(M_{Z}^{2})$ is to evaluate Eq.\eqref{EQ:dispersion} making use of $e^+e^-$ annihilation data for $R(s)$ in the  resonance regions, and either use the PQCD prediction for $R(s)$ above these regions (see e.g. \cite{davier2011}), or make use of all the available $e^+e^-$ data and fill in the gaps using the PQCD prediction (see e.g. \cite{hagiwara2011, Actis}). Since the use of data is the primary source of uncertainty,
other analyses have attempted to reduce the dependence of $\Delta\alpha_{\text{HAD}}(M_{Z}^{2})$ on $e^+e^-$ data  by a variety of methods that place a greater emphasis on PQCD. One approach in this direction is to subtract a polynomial from the weight function in Eq.\eqref{EQ:dispersion} to reduce the impact of the data contribution. In order to compensate, this polynomial weighted integral is added to the right hand side of Eq.\eqref{EQ:dispersion} and evaluated in PQCD (plus non-perturbative corrections given in the framework of the Operator Product Expansion) using a circular contour integral (see e.g. \cite{groote1998,kuhn1998}). Another approach is to first calculate $\Delta\alpha_{\text{HAD}}(-s_0)$  ($s_0>0$ with $s_0$  large enough for PQCD to be valid), whose weight function deemphasizes the low-energy region. Subsequently, $\Delta\alpha_{\text{HAD}}(-s_0)$ is  run to $\Delta\alpha_{\text{HAD}}(M_{Z}^2)$ using the PQCD prediction of the Adler function \cite{jegerlehner2008}.\\ 
The purpose of this paper is two-fold. First, to calculate the complete heavy quark (charm, bottom, and top) contributions to $\Delta\alpha_{\text{HAD}}(M_{Z}^2)$ using only PQCD, which to our knowledge has not been done before. Interestingly, this will significantly reduce the total uncertainty in $\Delta\alpha_{\text{HAD}}(M_{Z}^{2})$, as the use of $e^+e^-$ data in the charm-quark  region leads to an error equivalent to that from the use of $e^+e^-$ data in the light-quark resonance region. Second, to show how existing Lattice QCD (LQCD) calculations involved in the evaluation of the hadronic contribution to $g-2$ of the muon can be used to calculate the light-quark contribution to $\Delta\alpha_{\text{HAD}}(M_{Z}^{2})$ entirely from theory. This will allow for the first model-independent determination of $\Delta\alpha_{\text{HAD}}(M_{Z}^{2})$ that makes no use at all of $e^+e^-$ cross-section data. At present, though, current precision of LQCD results do not allow this method to compete with the standard approach.\\ 
\begin{figure}
\centering
\def\svgwidth{0.8\columnwidth}
\includegraphics[height=0.8in, width=3.3in]{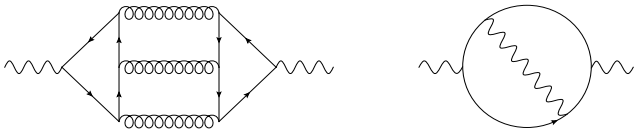}
\caption{{\protect\small{An example of a singlet three-gluon exchange diagram in QCD (left), and the lowest order QED correction to the vacuum polarization $\Pi(s)$ (right).}}}
\label{fig:qed}
\end{figure}
We begin by considering the heavy quark contribution to Eq.\eqref{eq:alpha1}, which can be  written as
\begin{equation}\label{eq:AlphaHeavy}
\Delta\alpha^{(f)}_{\text{HAD}}(M_{Z}^{2})=4\pi\alpha\left(\Pi^{(f)}(0)- Re \,[\Pi^{(f)}(M_{Z}^{2})]\right) \;,
\end{equation}
where $f=\{c,b,t\}$ are the heavy quark flavors, and in the sequel it should be understood that it is the real part of the correlators that enters in the time-like region.
It should be noticed that  one only needs knowledge of the correlator at $s=0$ and at $s=M_{Z}^{2}$. The latter scale is way above either the charm- or the bottom-quark pair production resonance region, so that one can safely use the high-energy expansion of the heavy quark correlator. This is known to $\mathcal{O}(\alpha_{s}^{3})$ (with partial results at $\mathcal{O}(\alpha_{s}^{4})$). In addition, $\Pi^{(f)}(0)$ has also been calculated in PQCD to $\mathcal{O}(\alpha_{s}^{3})$. The other key inputs are the recent high precision bottom- and charm-quark masses obtained from LQCD \cite{lattice} in the $\overline{\text{MS}}$-scheme. This scheme will be used  here in all  PQCD calculations. Prior to these LQCD determinations, the charm- and bottom-quark masses were obtained using  $e^+e^-$ data, a procedure we wish to avoid as we aim at an entirely theoretical determination of $\Delta\alpha^{(f)}_{\text{HAD}}(M_{Z}^{2})$. As explained later, it turns out that Eq.\eqref{eq:AlphaHeavy} is problematic for the charm-quark contribution (but not for the bottom- or the top-quark). The reason being its strong dependence  on the renormalization scale, which must be the same for both $\Pi^{(c)}(0)$ and $\Pi^{(c)}(M_{Z}^2)$. Therefore, we introduce two additional approaches in the charm-quark sector which are significantly less sensitive to this problem. The first is inspired by the Adler function approach of \cite{jegerlehner2008}, to wit. We note that
\begin{equation}\label{eq:ADLER}
\frac{d}{ds}\Delta\alpha^{(c)}_{\text{HAD}}(s)= - 4\pi \alpha \frac{d}{ds}\Pi^{(c)}(s)=\frac{\alpha}{3 \,\pi} \frac{D^{(c)}(s)}{s} \;, 
\end{equation}
where the real part is understood, and  $D^{(c)}(s)$ is the Adler function in the charm-quark channel. Integrating Eq.\eqref{eq:ADLER} gives
\begin{eqnarray}
\Delta\alpha^{(c)}_{\text{HAD}}(M^{2}_{z})&\equiv&\bigl[\Delta\alpha^{(c)}_{\text{HAD}}(M^{2}_{z})-\Delta\alpha^{(c)}_{\text{HAD}}(s_0)\bigr]\nonumber\\
&+&\Delta\alpha^{(c)}_{\text{HAD}}(s_0)=\frac{\alpha}{3\pi} \int^{M_{Z}^{2}}_{s_0}\frac{D^{(c)}(s)}{s}ds
\nonumber\\
&+& 4 \pi \alpha \left(\Pi^{(c)}(0)-\Pi^{(c)}(s_0)\right) \;.\label{eq:adler}
\end{eqnarray}
We choose $s_0$ large enough so that PQCD is valid, but still $s_0\ll M_{Z}^{2}$. One can then use one scale for the second term on the right hand side above, whilst another scale for integrating over the Adler function (one could also use a running scale, e.g. $\mu^2=s$). The second, but similar approach, is to use Cauchy's residue theorem to rewrite the dispersion relation Eq.\eqref{EQ:dispersion} to obtain 
\begin{eqnarray}
&&\Delta\alpha^{(c)}_{\text{HAD}}(M^{2}_{z})=4\, \alpha  \,M_{Z}^{2}\Bigl[ \frac{i}{2}\oint_{|s|=s_0}ds\,\frac{\Pi^{(c)}(s)}{s(M_{Z}^{2}-s)}\nonumber \\
&&+\pi \,\frac{\Pi^{(c)}(0)}{M_Z^2}\Bigr]+\frac{\alpha \, M_{Z}^{2}}{3 \,\pi}P\int^{\infty}_{s_0}\frac{R^{(c)}(s)}{s(M_{Z}^{2}-s)}ds \;, \label{eq:FESR}
\end{eqnarray}
which is only valid for $s_0<M_{Z}^{2}$. As usual, $s_0$ will be taken large enough so that PQCD is valid. Once again, this allows  for the use of more than one scale appropriate for the different regions. In addition, $R^{(c)}(s)$ is known partially up to $\mathcal{O}(\alpha_{s}^{4})$. The final virtue of this approach is that it will allow a careful region-by-region comparison  with the standard approach based on Eq.\eqref{EQ:dispersion}. The sum rule Eq.\eqref{eq:FESR} is very similar to the FESR used in precision charm- and bottom-quark mass determinations employing experimental data on $R(s)$ \cite{bodenstein2011,bodenstein2012}. Hence, to determine $\Delta\alpha^{(f)}_{\text{HAD}}(M^{2}_{z})$  entirely from theory it is essential to use a non-QCD sum rule determination of the charm- and bottom-quark masses, such as e.g. that from LQCD. The procedure just outlined for determining $\Delta\alpha^{(f)}_{\text{HAD}}(M^{2}_{z})$ is not necessary for the bottom- and top-quark counterparts, as Eq.\eqref{eq:AlphaHeavy} for $f=b,t$ gives results that are essentially renormalization scale independent. However, we have checked that the FESR and the Adler function approaches give the same result as that using Eq.\eqref{eq:AlphaHeavy} for the charm-quark contribution, although the latter has a much larger error.
\begin{figure}
\includegraphics[height=2.8in, width=3.6in]{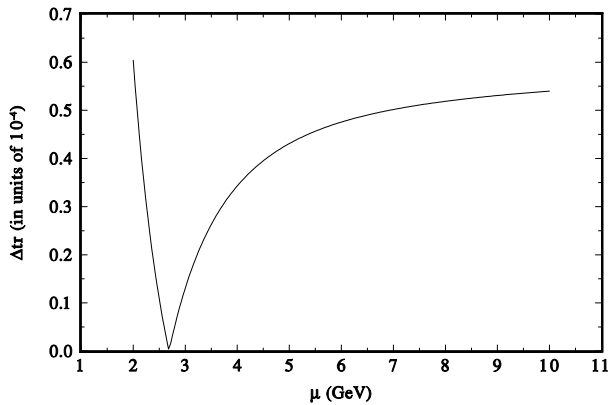}
\centering
\def\svgwidth{0.8\columnwidth}
\caption{{\protect\small{The difference between the order $\mathcal{O}(\alpha_{s}^{3})$ and the order $\mathcal{O}(\alpha_{s}^{2})$ results for $4\pi\alpha(\Pi^{(c)}(0)-\text{Re}[\Pi^{(c)}(M_{Z}^{2})])$, i.e. the truncation error $\Delta \text{tr}$, as a function of the renormalization scale $\mu$.  }}}\label{fig:convergence}
\end{figure}
\section{VECTOR CURRENT CORRELATOR IN QCD}
We provide in this section a summary of the available theoretical information on the  vector current correlator in QCD.  The flavor $f$-quark current correlator can be split as 
\begin{equation}\label{eq:total}
\Pi^{(f)}(s)=\Pi^{(f)}_{\text{PQCD}}(s)+\Pi^{(f)}_{\text{NP}}(s)+\Pi^{(f)}_{\text{QED}}(s) \;,
\end{equation}
where $f\in\{uds,c,b,t\}$, $\Pi^{(f)}_{\text{PQCD}}$ is the PQCD contribution, $\Pi^{(f)}_{\text{NP}}$ is the contribution from non-perturbative power corrections given in the framework of the Operator Product Expansion (OPE), and $\Pi^{(f)}_{\text{QED}}$ are QED corrections as shown in Fig. \ref{fig:qed}. 
The dominant contribution to Eq.\eqref{eq:total} is  the perturbative part $\Pi^{(f)}_{\text{PQCD}}$. In the massless case, appropriate for the $u,d,s$ quarks, the high energy limit of $\Pi^{(f)}_{\text{PQCD}}$ is known exactly up to order $\mathcal{O}(\alpha_{s}^{3})$, and up to a real constant to order $\mathcal{O}(\alpha_{s}^{4})$ (the full result is given in computer readable form in \cite{QCD0}). In the heavy-quark case, though, one needs both the low- and the high-energy expansions of the correlator. In the low-energy limit,
 with a single heavy quark and $n_f$ active flavors, the vector correlator can be written as
\begin{equation}
\Pi_f(s)=\frac{3 Q^{2}_{f}}{16\, \pi^2}\sum_{i=0}^{\infty}\bar{C}_i  z^i \;, \label{eq:LE}
\end{equation}
where $z\equiv s/(4\bar{m}_{f}^{2})$, and $\bar{m}_{f}$ is the mass of the quark of flavor-$f$ in the $\overline{\text{MS}}$ scheme at the scale $\mu$. The coefficients $\bar{C}_0$ and $\bar{C}_1$ were determined up to $\mathcal{O}(\alpha_{s}^{3})$ in \cite{QCD1,QCD2},  $\bar{C}_2$ in \cite{C2}, and $\bar{C}_3$ in \cite{C3}. In the high-energy limit the heavy quark correlator is written as the  massless one with added quark-mass corrections
\begin{equation}
\Pi(s)=Q_{f}^{2}\sum_{n=0}^{\infty}\left(\frac{\alpha_s(\mu^2)}{\pi}\right)^n\Pi^{(n)}(s)\;,
\end{equation} 
where 
\begin{equation}
\Pi^{(n)}(s)=\sum_{i=0}^{\infty}\left(\frac{\bar{m}^{2}_{f}(\mu)}{s}\right)^i\Pi^{(n)}_i \;.
\end{equation}
\begin{figure}[ht]
\begin{center}
\includegraphics[height=3.0in, width=3.5in]{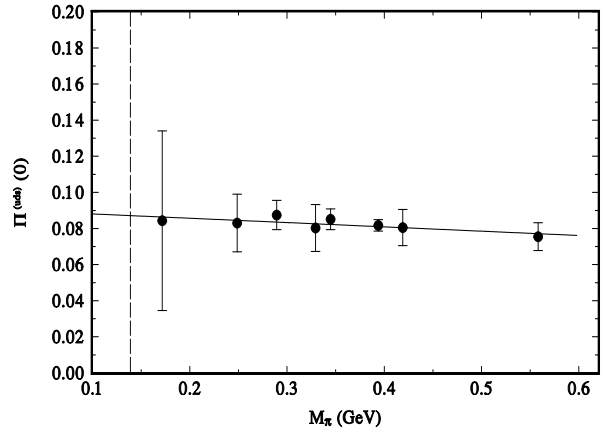}
\caption{{\protect\small{ The value of $\Pi^{(\text{uds})}(0)$ in LQCD in the $\overline{\text{MS}}$-bar scheme at $\mu=2\,\text{GeV}$
for different values of the pseudoscalar mass. Solid line is a linear fit extrapolation to the physical pion mass (indicated by the vertical dashed line). The errors from the LQCD fit parameters are added in quadrature.}}}\label{Fig:data}
\end{center}
\end{figure}
\begin{table*}
\begin{ruledtabular}
\begin{tabular}{lccccccc}
 & \multicolumn{6}{c}{Contributions to $\Delta\alpha_{\text{HAD}}$ (${\mbox{in units of}}\, 10^{-4})$} \\
\cline{2-7}
\noalign{\smallskip}
	 																			&  $[0-3.7\,\text{GeV}]_{uds}$			&	$[3.7-9.3\,\text{GeV}]_{udsc}$			& $[9.3-40\,\text{GeV}]_{udscb}$	& $[>40\,\text{GeV}]_{udscb} $ & $[>40\,\text{GeV}]_t$    & Total  \\
\hline
\noalign{\smallskip}
Standard approach \cite{davier2011}		&	$79.29\pm 0.69$		&	$60.21\pm 0.51$	& $93.50\pm 0.16$		& $42.70\pm 0.06$			&	$-0.72\pm 0.01$			&	$275.0\pm 1.0$			\\ \\
This work	& 	            $79.39\pm 0.68$		&	$60.46\pm 0.33$		& $93.82\pm 0.14$		& $42.76\pm 0.06$		&	$-0.76\pm 0.03$			& $275.7\pm 0.8$			  \\     
\end{tabular}
\caption{The contributions to $\Delta\alpha_{\text{HAD}}$ from different regions using either Eq.\eqref{EQ:dispersion} (standard approach), or Eq.\eqref{eq:FESR} (this work) which requires data only for $s < 1.8 \,\mbox{GeV}^2$ and PQCD above this energy. Our results in columns three and four are obtained using Eq.\eqref{eq:FESR} also for the bottom-quark contribution, with $\sqrt{s_0} = 40 \, {\mbox{GeV}}$.
The total error takes into account the correlations of the uncertainties from the different regions.}\label{Tab:results}
\end{ruledtabular}
\end{table*}

The terms $\Pi^{(0)}$ and $\Pi^{(1)}$ above are known exactly, whilst $\Pi^{(2)}$ is for all practical purposes also known exactly, i.e. the mass corrections up to $\mathcal{O}(m^{60})$ are given in \cite{QCD2,Pi2a,Pi2b,QCD3}. At order $\mathcal{O}(\alpha_{s}^{3})$, there is only partial information on the mass corrections. The functions $\Pi^{(3)}_0$ and $\Pi^{(3)}_1$ are known exactly \cite{QCD4}, whilst the logarithmic terms of $\Pi^{(3)}_1$ are given in \cite{QCD5}. The constant term has been estimated using Pade approximants \cite{QCD6}, but it was found to be negligible. At order $\mathcal{O}(\alpha_{s}^{4})$  the logarithmic terms in $\Pi^{(4)}_0$ and $\Pi^{(4)}_1$ are known \cite{QCD7,QCD8}, but not the constant terms. Hence, for the heavy quarks no $\mathcal{O}(\alpha_{s}^{4})$ terms will be included in any contour integral. However, $\Pi^{(4)}_0$ and $\Pi^{(4)}_1$ will be included when integrating over  $\text{Im}(\Pi(s))$, as the unknown constant terms do not contribute in this case. At $\mathcal{O}(\alpha_{s}^{3})$, there are singlet diagrams contributing for the first time (Fig. \ref{fig:qed}). The logarithmic terms are known in the high energy case \cite{QCD5}, and we will use these in Eq.\eqref{EQ:dispersion} to estimate the singlet contribution. An example of a lowest order contribution to $\Pi^{(f)}_{\text{QED}}(s)$ is shown in Fig. \ref{fig:qed}, after substituting gluons by photons.  Finally, the leading-order non-perturbative contribution to $\Pi^{(f)}_{\text{NP}}(s)$ for heavy quarks is from the gluon condensate $\left<(\alpha_s/\pi)G^2\right>$. This has been determined  from data on $\tau$-decays  \cite{G2tau}, and with a very large uncertainty  from data on $e^+ e^-$ annihilation into hadrons \cite{G2R}. The conservative value $\left<(\alpha_s/\pi)G^2\right>=(0.006\pm 0.012)\,\text{GeV}^4$ will be used in the sequel. To estimate the error arising from the incomplete knowledge of the correlator in PQCD (truncation error), we  take the difference between the $\mathcal{O}(\alpha_{s}^n)$ and $\mathcal{O}(\alpha_{s}^{n-1})$ results, where $n$ is the highest available order. 
We will check this by also varying the scale $\mu$. Finally, many errors will be 100\% correlated or anti-correlated between different regions, such as e.g. the error in $\alpha_s$.\\
As input, the PDG value of the Z-mass will be used, i.e. $M_Z=91.1876(21)$ \cite{PDG}. For the strong coupling we use the result from the determination of Davier {\it et al.} \cite{davier2011} 
\begin{equation}\label{eq:coupling}
\alpha_{s}(M_{Z}^{2})=0.1193(28) \;,
\end{equation}
in order to facilitate the comparison of the final result for $\Delta\alpha_{\text{HAD}}(M_{Z}^{2})$. This value has a far more conservative error than the PDG value of $\alpha_{s}(M_{Z}^{2})=0.1184(7)$ \cite{PDG}. However, this will hardly matter as  our results turn out to be largely independent of the strong coupling. For the charm- and bottom-quark masses, we use the most recent LQCD determination \cite{lattice}, $\bar{m}_{c}^{(4)}(3\,\text{GeV})=0.986(6)\,\text{GeV}$ and $\bar{m}_{b}^{(5)}(10\,\text{GeV})=3.617(25)\,\text{GeV}$. These values are in very good agreement with QCD sum rule determinations \cite{bodenstein2011,bodenstein2012,kuhn2007}. For the top-quark mass, we use $\bar{m}_{t}^{(5)}(\bar{m}_{t})=160.0(3.5)\,\text{GeV}$ \cite{topmass}. If one employs either the Adler function or the FESR approach,  the coupling and quark masses  need to be run across flavor thresholds.
This will be done using  the Mathematica program \texttt{RunDec} \cite{rundec}.\\
\section{CHARM-QUARK CONTRIBUTION}
We consider first the evaluation of $\Delta\alpha_{\text{HAD}}^{(c)}(M_{Z}^{2})$ directly from  Eq.\eqref{eq:alpha1} . As mentioned previously, there is a problem with the strong dependence of $\Delta\alpha_{\text{HAD}}^{(c)}(M_{Z}^{2})$ on the renormalization scale $\mu$. In fact, if one varies $\mu$ in the interval $\mu = 2\,\text{GeV} - M_{Z}$ then $\Delta\alpha^{(c)}_{\text{HAD}}(M_{Z}^{2})$ changes by $0.77\times 10^{-4}$, which is a large variation in the context of the current precision of $\Delta\alpha_{\text{HAD}}(M_{Z}^{2})$. We  choose instead a value of  $\mu$ leading to a good  convergence of the perturbative series. Figure \ref{fig:convergence} shows the difference between the order $\mathcal{O}(\alpha_{s}^{3})$ and  the order $\mathcal{O}(\alpha_{s}^{2})$ results for $4\pi\alpha(\Pi^{(c)}(0)-\text{Re}[\Pi^{(c)}(M_{Z}^{2})])$, i.e. the truncation error $\Delta \text{tr}$, as a function of the renormalization scale $\mu$. A value of
 $\mu \simeq 3\,\text{GeV}$ ensures good convergence. Using $n_f=5$, this leads to 
\begin{eqnarray}
\Delta\alpha^{(c)}_{\text{HAD}}(M_{Z}^{2})&=&4\pi\alpha\left(\Pi^{(c)}(0)-  \,[\Pi^{(c)}(M_{Z}^{2})]\right)\nonumber\\
&=& (79.19\pm 0.13_{\Delta \text{tr}}\pm 0.03_{\Delta \alpha_s} \nonumber\\
&\pm&  0.01_{\Delta\langle G^2\rangle}  \pm 0.11_{\Delta\bar{m}_c})\times 10^{-4},
\end{eqnarray}
where the errors are due to truncation ($\Delta \text{tr}$), and to uncertainties in $\alpha_s$ ($\Delta \alpha_s$), in the gluon condensate ($\Delta \langle G^2 \rangle$), and in the charm-quark mass ($\Delta \bar{m}_c$).
Interestingly, this value corresponds to  a global minimum of $\Delta\alpha^{(c)}_{\text{HAD}}(M_{Z}^{2})$ as a function of $\mu$. In any case, given this strong $\mu$-dependence we shall not use this method to determine the charm-quark contribution. 
We consider instead the Adler function approach using Eq.\eqref{eq:adler} which involves two terms, a high energy and a low energy contribution. Starting with the low energy part, and using $n_f=4$, $s_0=(9.3\,\text{GeV})^2$ (which is below the bottom-quark threshold), and an initial value $\mu=5\,\text{GeV}$ (to be made to vary in a wide range later), we find 
\begin{eqnarray}
&&4 \,\pi \,\alpha \,(\Pi^{(c)}(0)- \,\Pi^{(c)}(s_0))=(29.57\pm 0.25_{\Delta \text{tr}}\nonumber\\
&&\pm \,0.05_{\Delta \alpha_s} \pm\, 0.01_{\Delta\langle G^2\rangle}  \pm \, 0.12_{\Delta\bar{m}_c})\times 10^{-4}\;. \label{eq:adlercharm1}
\end{eqnarray}
Allowing now $\mu$ to vary in the wide range $\mu = (2\,- 9.3)\,\text{GeV}$, results in a variation of the central value above in the range  $[29.31 - 29.62]\times 10^{-4}$, which is within the truncation error. For the high energy term, we use $n_f=5$ and an initial value $\mu=\frac{1}{2}M_{Z}^{2}$ to obtain
\begin{eqnarray}
\frac{\alpha}{3\pi} \int^{M_{Z}^{2}}_{s_0}\frac{D^{(c)}(s)}{s}ds &=&(49.91\pm 0.05_{\Delta \text{tr}} \pm \, 0.04_{\Delta \alpha_s} \nonumber\\
&\mp& 0.01_{\Delta\bar{m}_c})\times 10^{-4} \;. \label{eq:adlercharm2}
\end{eqnarray}
Varying  $\mu$ in the interval $\mu = 9.3\,\text{GeV} - M_Z$, produces a variation of the central value above in the range  $[49.82 - 49.95] \times 10^{-4}$. Adding both contributions, Eqs.\eqref{eq:adlercharm1} and \eqref{eq:adlercharm2}, the total result from this approach is 
\begin{eqnarray}
\Delta\alpha^{(c)}_{\text{HAD}}(M_{Z}^{2})&=&(79.49\pm 0.30_{\Delta \text{tr}}\pm 0.09_{\Delta \alpha_s}\nonumber\\
&\pm& 0.01_{\Delta\langle G^2\rangle}  \pm 0.11_{\Delta\bar{m}_c})\times 10^{-4} \;,
\end{eqnarray}
where we notice that the errors $\Delta\bar{m}_c$ in Eqs. \eqref{eq:adlercharm1} and \eqref{eq:adlercharm2} are anti-correlated.
The third method consists in using the FESR Eq.\eqref{eq:FESR}. The values $n_f=4$, and $\mu=5\,\text{GeV}$, have been used both in the contour integral, with $s_0 = (9.3\,\text{GeV})^2$, as well as in the residue $\Pi^{(c)}(0)$. For the integral involving $R^{(c)}(s)$ we use $n_f=5$ (the contribution above the top threshold is not numerically important), and obtain
\begin{eqnarray}\label{eq:charmFESR}
\Delta\alpha^{(c)}_{\text{HAD}}(M_{Z}^{2})&&=(79.34\pm 0.26_{\Delta \text{tr}}\pm 0.04_{\Delta \alpha_s}\nonumber\\
&& \pm 0.01_{\Delta\langle G^2\rangle}  \pm 0.11_{\Delta\bar{m}_c})\times 10^{-4} \;.
\end{eqnarray}
Varying $\mu$ in the respective regions as done previously produces changes in the central value of $\Delta\alpha^{(c)}_{\text{HAD}}(M_{Z}^{2})$  well within the truncation error. This result will be adopted for the charm-quark contribution as it is has a much smaller uncertainty due to $\alpha_s$, and a slightly smaller truncation error. It should be mentioned that at $s=0$ the low energy expansion, Eq.\eqref{eq:LE}, is a well convergent power series expansion in the strong coupling. In the vicinity of $s=0$, Eq.\eqref{eq:LE} also converges well, even if the charm-quark is at the borderline between light and heavy quarks. This is a standard procedure in the determinations of the charm-quark mass from QCD sum rules or from LQCD.
\\ 
\section{BOTTOM- AND TOP-QUARK CONTRIBUTIONS}
In this case all three methods give essentially the same result for $\Delta\alpha^{(b)}_{\text{HAD}}(M_{Z}^{2})$. Using e.g. Eq.\eqref{eq:alpha1} with $n_f=5$ and $\mu=10\,\text{GeV}$   gives
\begin{eqnarray}\label{eq:beauty}
\Delta\alpha^{(b)}_{\text{HAD}}(M_{Z}^{2})&=& 4\pi\alpha\left(\Pi^{(b)}(0)-  \,[\Pi^{(b)}(M_{Z}^{2})]\right)\nonumber\\
&=&(12.79\pm 0.06_{\Delta \text{tr}}\pm 0.009_{\Delta \alpha_s} \nonumber\\
 &\pm& 0.03_{\Delta\bar{m}_b})\times 10^{-4}\;.
\end{eqnarray}
Varying $\mu$ in the very wide range $\mu = 10\,\text{GeV} - 10 M_Z$  only changes this result  by $0.04\times 10^{-4}$, which shows a remarkable scale independence. \\ 
For the top quark, one can use the low energy expansion to calculate both $\Pi^{(t)}(0)$ and  $\Pi^{(t)}(M_{Z}^{2})$. Up to $\mathcal{O}(\alpha_s)$ the full analytic correlator is known. We have verified that there is no appreciable difference between results using the low-energy expansion of the correlator or using the full expression up to this order to determine $\Pi^{(t)}(M_{Z}^{2})$. At higher orders, one can reconstruct the full analytic behavior of the correlator using Pade approximants, as done in  \cite{QCD6} at order $\mathcal{O}(\alpha^{3}_{s})$. Using these results we find that it is perfectly safe to use the low energy expansion of the correlator. With $\mu=\bar{m}_{t}$ and $n_f=6$, we find 
\begin{eqnarray} \label{eq:truetop}
\Delta\alpha^{(t)}_{\text{HAD}}(M_{Z}^{2})&=& 4 \pi \alpha\left(\Pi^{(t)}(0)- \, \Pi^{(t)}(M_{Z}^{2})\right)\nonumber\\
&=&(-0.76\pm 0.03_{\Delta\bar{m}_t})\times 10^{-4} \;,
\end{eqnarray}
where only the uncertainty in the top-quark mass produces a non-negligible uncertainty in $\Delta\alpha^{(t)}_{\text{HAD}}(M_{Z}^{2})$. 
\begin{table*}[ht!]
\begin{ruledtabular}
\begin{tabular}{llc}

\noalign{\smallskip}
	 		 Authors & $\Delta\alpha^{(5)}_{\text{HAD}}(M_{Z}^{2}) ({\mbox{in units of}}\, 10^{-4})$   & Method\\																	  
\hline
\noalign{\smallskip}
Groote \emph{et al.} (1998) \cite{groote1998}					& 277.6(4.1)						& PQCD driven (using polynomial weights and global duality) \\
K\"uhn \emph{et al.} (1998) \cite{kuhn1998}						& 277.5(1.7)						& PQCD driven (using polynomial weights and global duality) \\
Burkhardt \emph{et al.} (2011) \cite{burkhardt2011}		& 275.0(3.3)						& Data driven\\
Troc\'oniz \emph{et al.} (2005) \cite{troconiz2005}		& 274.9(1.2)					  & Data driven   \\
Jegerlehner (2008) \cite{jegerlehner2008}							&	275.94(2.19)					& Data driven \\
Jegerlehner (2011) \cite{jegerlehner2011}							& 274.98(1.35)					& Adler function approach\\
Hagiwara \emph{et al.} (2011) \cite{hagiwara2011}			&	276.26(1.38)	 				& Data driven\\
Davier \emph{et al.} (2011) \cite{davier2011}					&	275.7(1.0)	 					& PQCD in range $1.8<\sqrt{s}<3.7\,\text{GeV}$, and for $\sqrt{s}>5\,\text{GeV}$, \\ 
									& & otherwise data.\\
This work 																						&	276.5(0.8)	 					& Data for $\sqrt{s}<1.8\,\text{GeV}$, and PQCD for $\sqrt{s}>1.8\,\text{GeV}$. \\
This work																						&	$\sim 273$		 		    & LQCD + PQCD \\	 					                     
\end{tabular}
\caption{Some five-flavor results of previous analyses of $\Delta\alpha_{\text{HAD}}(M_{Z}^{2})$   together with our determinations. Only  the latest available results from each collaboration are quoted. No error is given in the last line (LQCD + PQCD) in view of the large uncertainties from LQCD.}\label{Tab:resultsOther}
\end{ruledtabular}
\end{table*}
\section{LIGHT-QUARK CONTRIBUTION}
In contrast to the heavy quark contributions, one cannot use PQCD to determine the light-quark correlator at low energies. There are two approaches to achieve this, i.e. using $e^+e^-$ data for $R(s)$, or LQCD determinations of $\Pi(s)$ (in the space-like region), with both being used in the sequel. For the $e^+e^-$ data approach, and  below the onset of PQCD, we use the integrated result of  \cite{davier2011} to avoid the complicated task of dealing with the vast amount of $e^+e^-$ data available (for a recent independent analysis see \cite{G2R}). The  result of \cite{davier2011}, integrated up to the PQCD threshold $\sqrt{s}=1.8\,\text{GeV}$ is 
\begin{align} \label{eq:light}
\frac{\alpha M_{Z}^{2}}{3 \pi} &  \int^{(1.8 \text{GeV})^2}_{0}\frac{R^{(uds)}_{\text{data}}(s)}{s(M_{Z}^{2}-s)}ds \nonumber\\
&= (55.02\pm 0.66)\times 10^{-4} \;.
\end{align}
Above the PQCD threshold, and using the massless order $\mathcal{O}(\alpha_{s}^{4})$ PQCD expression for $R^{(uds)}(s)$ we find
\begin{eqnarray}
&&\frac{\alpha M_{Z}^{2}}{3\pi}\int^{\infty}_{(1.8\,\text{GeV})^2}\frac{R^{(uds)}_{\text{PQCD}}(s)}{s(M_{Z}^{2}-s)}ds \nonumber\\
&=& (129.26\pm 0.16_{\Delta \text{tr}}
\pm 0.29_{\Delta \alpha_s})\times 10^{-4}, 
\end{eqnarray}
which added to Eq.\eqref{eq:light} gives the total light-quark contribution
\begin{eqnarray}\label{eq:totaluds}
\Delta\alpha^{(uds)}_{\text{HAD}}(M_{Z}^{2})&=&(184.28\pm 0.66_{\text{data}}\pm 0.16_{\Delta \text{tr}}\nonumber\\
&\pm& 0.29_{\Delta \alpha_s})\times 10^{-4}\;.
\end{eqnarray}
A large effort is currently underway to determine $\Pi^{(uds)}(s)$ in the space-like region using LQCD.
A key aim is to  provide a first-principles determination of the hadronic contribution to the $g-2$ of the muon \cite{Blum,Aubin,boyle2011,renner2011,wittig2012}. Here we describe two methods for obtaining $\Delta\alpha^{(uds)}_{\text{HAD}}(M_{Z}^{2})$, entirely from theory, from a combination of LQCD and PQCD. This is inspired by \cite{g-2BOD}, where an entirely theoretical determination of $g-2$ was proposed.
It must be emphasized, though,  that LQCD results are currently not precise enough to compete with the $e^+e^-$ approach. For instance, one source of uncertainty arises from disconnected Feynman diagrams, which are currently not included in LQCD calculations, and which lead to an estimated 10\% systematic uncertainty \cite{boyle2011}.\\ 
The first method is based on the FESR Eq.\eqref{eq:FESR}, with $\Pi^{(uds)}(0)$  determined from LQCD, and the two integrals computed in PQCD. In LQCD it is not possible to  calculate directly $\Pi^{(uds)}(0)$. Instead, the correlator is computed for values very close to $s=0$, and then these results are fitted and extrapolated to the origin to obtain $\Pi_{uds}(0)$. For instance, the phenomenologically inspired fitting function used in \cite{boyle2011} is  of the form
\begin{equation}\label{eq:fit}
\Pi_{uds}(s)=A-\frac{F_{1}^{2}}{m_{1}^{2}-s}-\frac{F_{2}^{2}}{m_{2}^{2}-s} \;,
\end{equation} 
with fit parameters  given in \cite{boyle2011}. An alternative, model independent approach to extrapolating LQCD data is based on Pade approximants \cite{Aubin2}. This approach would be appropriate in future precision determinations based on improved LQCD data. An important observation is that neither $\Pi_{uds}(0)$ nor the contour integral in Eq.\eqref{eq:FESR} are observable quantities. Therefore, it is essential to compute both of these quantities in the same renormalization scheme, and at the same scale, so that the observable difference between $\Pi_{uds}(0)$ and the contour integral is scheme-independent. A problem arises because
PQCD schemes,  such as  $\overline{\text{MS}}$, are not easy to relate to LQCD renormalization schemes. The latter lead  to a prediction of $\Pi^{(uds)}(s)$ which differs from the  $\overline{\text{MS}}$ results by the constant $\Pi^{(uds)}(0)$, which is precisely what is needed in Eq.\eqref{eq:FESR}. The standard approach to fix this constant is to impose agreement between PQCD and  LQCD results  at some value $s = -s^*$ where PQCD is expected to be valid. This procedure would then allow for a determination of $\Pi^{(uds)}(0)$ from LQCD. Above $s \simeq - 2 \, {\mbox{GeV}}^2$ LQCD results already are in agreement with PQCD, so to determine $\Pi^{(uds)}(0)$ we choose $s^*= - 3.5\,\text{GeV}^2$ to be on the safe side, together with $s_0=(3.72\,\text{GeV})^2$ which corresponds to the onset of the charm-quark region, and 
$n_f=3$. The renormalization scale was varied in the wide range between the $\tau$-lepton mass and the  charm threshold, i.e.  $\mu=1.77 - 3.7\,\text{GeV}$. This produces a negligible change in  $\Delta\alpha^{(uds)}_{\text{HAD}}(M_{Z}^{2})$ of $0.05 \times 10^{-4}$.
There are eight available LQCD fits \cite{boyle2011} for different  pion masses, with each giving a result for $\Pi^{(uds)}(0)$ in need of extrapolation to the actual physical value of the pion mass. Using a simple linear extrapolation fit gives the results shown in Fig. \ref{Fig:data}, leading to
\begin{equation}
\Pi^{(uds)}(0)\sim 0.08758 \ \ (\mu=2\,\text{GeV}) \;.
\end{equation}
Using this value in Eq.\eqref{eq:FESR}, together with a PQCD evaluation of the integrals, gives
\begin{equation}\label{eq:uds}
\Delta\alpha^{(uds)}_{\text{HAD}}(M_{Z}^{2})\sim 181 \times 10^{-4} \;.
\end{equation}
No error is given above in view of the current uncertainties in LQCD.\\
The second method to determine $\Delta\alpha^{(uds)}_{\text{HAD}}(M_{Z}^{2})$ entirely from theory makes
use of the  Adler function. One important advantage over the FESR approach is that there is no need to enforce the matching between PQCD and LQCD at any point. In order to involve the Adler function we write
$\Delta\alpha^{(uds)}_{\text{HAD}}(M_{Z}^{2})$ as
\begin{widetext}
\begin{eqnarray} \label{eq:last}
\Delta\alpha^{(uds)}_{\text{HAD}}(M_{Z}^{2})&\equiv&\Delta\alpha^{(uds)}_{\text{HAD}}(-s_0)+\bigl[\Delta\alpha^{(uds)}_{\text{HAD}}(s_0)  
 - \Delta\alpha^{(uds)}_{\text{HAD}}(-s_0)\bigr]
+\bigl[\Delta\alpha^{(uds)}_{\text{HAD}}(M_z^2)-\Delta\alpha^{(uds)}_{\text{HAD}}(s_0)\bigr]\nonumber\\
&=& 4\pi\alpha \;\left[\Pi^{(uds)}_{\text{LQCD}}(0)-\Pi^{(uds)}_{\text{LQCD}}(-s_0)\right]
+\frac{\alpha}{3\pi} \int^{s_0}_{-s_0}\frac{D^{(uds)}_{\text{PQCD}}(s)}{s}ds 
+\frac{\alpha}{3\pi} \int^{M_{Z}^{2}}_{s_0}\frac{D^{(uds)}_{\text{PQCD}}(s)}{s}ds ,
\end{eqnarray}
\end{widetext}
where $s_0$ is large enough for PQCD be valid, and the real part of the expression in the last line is to be understood. Notice that the line integral in the interval $(-s_0, + s_0)$ is well defined. In fact, since $D(s)$ is an analytic function the integration was performed on a semi-circular contour  of radius $|s_0|$, avoiding the origin. Evaluating each of the three terms on the right hand side of Eq.\eqref{eq:last}, with $s_0 = s^* = - 3.5 \; {\mbox{GeV}}^2$,
we find
\begin{equation} \label{eq:LQCDuds}
4\pi\alpha \;\left[\Pi^{(uds)}_{\text{LQCD}}(0)-\Pi^{(uds)}_{\text{LQCD}}(-s_0)\right] \simeq 52.32 \, \times 10^{-4}\,,
\end{equation}
\begin{equation}
\frac{\alpha}{3\pi} \int^{s_0}_{-s_0}\frac{D^{(uds)}_{\text{PQCD}}(s)}{s}ds = 2.56 \, \times 10^{-4} \,,
\end{equation}
\begin{equation}
\frac{\alpha}{3\pi} \int^{M_{Z}^{2}}_{s_0}\frac{D^{(uds)}_{\text{PQCD}}(s)}{s}ds = 125.95 \, \times 10^{-4} \,,
\end{equation}
which add up to
\begin{equation} \label{eq:total2}
\Delta\alpha^{(uds)}_{\text{HAD}}(M_{Z}^{2}) = 181 \times 10^{-4}\,,
\end{equation}
as already given in Eq.\eqref{eq:uds}. The errors in the total PQCD contributions  are
\begin{align} \label{eq:errorPQCD}
\frac{\alpha}{3\pi}& \int^{s_0}_{-s_0}\frac{D^{(uds)}_{\text{PQCD}}(s)}{s}ds +
\frac{\alpha}{3\pi} \int^{M_{Z}^{2}}_{s_0}\frac{D^{(uds)}_{\text{PQCD}}(s)}{s}ds \nonumber\\
&= (128.5 \pm 0.2_{\Delta \text{tr}} \pm 0.3 _{\Delta \alpha_s}) \times 10^{-4} \,.
\end{align}
The contribution of the gluon condensate is at the level of one order of magnitude smaller than the uncertainty in $\alpha_s$, hence it can be neglected.
Once LQCD determinations of Eq.\eqref{eq:LQCDuds} achieve enough accuracy, it would become possible to determine $\Delta\alpha_{\text{HAD}}(M_{Z}^{2})$ entirely from theory, after adding to the LQCD light-quark contribution the heavy-quark results Eqs.\eqref{eq:charmFESR}, \eqref{eq:beauty}, and \eqref{eq:truetop}.
\section{CONCLUSIONS}
Adding up all of the contributions, i.e. Eqs.\eqref{eq:charmFESR}, \eqref{eq:beauty}, \eqref{eq:truetop}, and \eqref{eq:totaluds} , the final result for $\Delta\alpha_{\text{HAD}}(M_{Z}^{2})$ is
\begin{eqnarray}\label{eq:finalf}
\Delta\alpha_{\text{HAD}}(M_{Z}^{2})&=&
\left(275.7 \pm 0.66_{\Delta \text{data}} \pm 0.44_{\Delta \text{tr}} \pm 0.26_{\Delta \alpha}   \right.\nonumber \\
&\pm& \left. 0.11_{\Delta \bar{m}_c}\right)\times 10^{-4} \nonumber\\
&=&
(275.7\pm 0.8)\times 10^{-4} \;,
\end{eqnarray}
where $n_f = 6$ has been used, and the uncertainties due to the bottom- and the top-quark masses, and due to the gluon condensate, are negligible. This result can be compared with $\Delta\alpha_{\text{HAD}}(M_{Z}^{2})= (275.0\pm 1.0)\times 10^{-4}$ from \cite{davier2011} (for $n_f=6$)(other five quark flavor results are listed in Table II). The primary reason for this $20\%$ reduction in uncertainty is our PQCD calculation of the contribution of the charm-quark resonance region, which is given in Table \ref{Tab:results}.\\

We comment in closing on the relation between $\Delta\alpha_{\text{HAD}}(M_{Z}^{2})$ and the value of the Higgs mass. To begin with, as is well known both $\Delta\alpha_{\text{HAD}}(M_{Z}^{2})$ and the QCD strong coupling $\alpha_s(M_Z^2)$ enter into the global SM fit to electroweak precision data. At first sight one could suspect that the central role played by PQCD in our approach could lead to a stronger correlation  between $\Delta\alpha_{\text{HAD}}(M_{Z}^{2})$ and $\alpha_s(M_Z^2)$. However, this is not the case. In fact, quite the contrary, i.e. in comparison with the standard approach our FESR method reduces the overall dependence of  $\Delta\alpha_{\text{HAD}}(M_{Z}^{2})$ on $\alpha_s(M_Z^2)$. This happens because the $\alpha_s(M_Z)$ dependence of the combination of contour integral plus Cauchy residue in Eq.\eqref{eq:FESR} is anti-correlated with the $\alpha_s(M_Z)$ dependence of the integral involving $R(s)$. Hence, there is some cancellation of the $\alpha_s(M_Z)$ dependence, which does not take place in the standard approach which uses data to determine the heavy-quark resonance contribution. Quantitatively, the functional dependence of the central value of $\Delta\alpha_{\text{HAD}}(M_{Z}^{2})$ on the value of $\alpha_s(M_Z)$ in the standard approach \cite{davier2011} is approximately  $132.1 [\alpha_s(M_Z) - 0.1193] \times 10^{-4}$, while in our method it becomes approximately $107.7 [\alpha_s(M_Z) - 0.1193] \times 10^{-4}$.\\
The correlation between $\Delta\alpha_{\text{HAD}}^{(5)}(M_{Z}^{2})$ and the logarithm of the Higgs mass, $\ln{M_H}$, was found in \cite{HM} to be $-0.395$ (for an earlier determination see \cite{HM0}). Using our result for $\Delta\alpha_{\text{HAD}}^{(5)}(M_{Z}^{2})$ in Table II would lead to a Higgs mass $M_H \simeq 87 \, {\mbox{GeV}}$, somewhat lower than the value from \cite{davier2011} $M_H \simeq 91^{+30}_{-23} \, {\mbox{GeV}}$, thus increasing the tension between a possible Higgs of mas $M_H \simeq 126 \, {\mbox{GeV}}$, and the fitted Higgs mass.

\section{ACKNOWLEDGMENTS}
This work was supported in part by the National Research Foundation (South Africa) and by the Alexander von Humboldt Foundation (Germany). The authors thank A. Denig, M. Fritsch, and F. Jegerlehner for helpful discussions.

\end{document}